# Mechanical investigations of composite cathode degradation in all-solid-state-batteries

Shafee Farzanian [1], Imtiaz Shozib [1], Nikhil Sivadas [2], Valentina Lacivita [2], Yan Wang [2]*, Qingsong Howard Tu [1]*

1. Department of Mechanical Engineering, Rochester Institute of Technology, Rochester, NY 14623, USA

2. Advanced Materials Lab, Samsung Advanced Institute of Technology-America, Samsung Semiconductor Inc., Cambridge, MA 02138, USA

*Corresponding author: Qingsong Howard Tu (Email: howard.tu@rit.edu),

Yan Eric Wang (Email: eric.wangyan@samsung.com)

Lead contact: Qingsong Howard Tu (Email: howard.tu@rit.edu)

**Abstract**

Despite ongoing efforts aimed at increasing energy density in all-solid-state-batteries, the optimal composite cathode morphology, which requires minimal volume change, small void development, and good interfacial contact, remains a significant concern within the community. In this work, we focus on the theoretical investigation of the above-mentioned mechanical defects in the composite cathode during electrochemical cycling. It is demonstrated that these mechanical defects are highly dependent on the SE material properties, the external stack pressure and the cathode active material (CAM) loading. The following conclusions are highlighted in this study:

**(1)** Higher CAM loading (>50 vol. %) causes an increase in mechanical defects, including large cathode volume change (>5%), contact loss (50%) and porosity (>1%).

**(2)** High external stack pressure up to 7MPa reduces mechanical defects while preventing internal fracture in the cathode.

**(3)** Soft SE materials with small Young's modulus (<10GPa) and low hardness (<2GPa) can significantly minimize these mechanical defects during cycling.

**(4)** A design strategy is proposed for high CAM loading with minimal mechanical defects when different SE materials are utilized in the composite cathode, including oxide-type SE, sulfide-type SE, and halide-type SE.

The research provides specific guidelines to optimize the composite cathode in terms of mechanical properties. These guidelines broaden the design approach towards improving the performance of SSB, by highlighting the importance of considering the mechanical properties of battery materials.

# Introduction

All-solid-state batteries (ASSBs) have emerged as promising alternatives to conventional lithium-ion batteries (LIBs) due to their higher energy densities, longer service lives and safer properties [1]. Despite tremendous efforts made toward the development of ASSBs, several challenges continue to impede the technology's full potential for commercial applications [2, 3]. These challenges include the growth of dendrites from the Li-metal anode through the solid electrolyte (SE) [4], interfacial issues such as chemical reactions between the electrodes and the SE [5], cell volume variations during cycling [6, 7], and microstructural deterioration in the composite cathode [8]. Although solutions have been proposed to overcome many of these challenges, the significant mechanical responses generated in the composite cathode remain a significant concern. These responses originate from inevitable lattice expansion-contraction of Cathode Active Materials (CAM) during (de)lithiation at atomic level [9], up to the meso-macro level including the cathode volume change [10], porosity [11], and fractures [12].

The composite cathode in SSBs is composed of CAM, SE, and carbon, and is typically made through combining these three components and pressing them with one layer of bulk separator and one layer of anode [5, 13, 14, 15, 16]. During charging and discharging cycles, the cyclic electro-chemo-mechanical loads cause the expansion and contraction of CAM particles, which generates considerable internal strain within the composite cathode. This strain leads to interfacial contact loss, porosity development, and permanent volume change in the SSBs. The mechanical characteristics of materials used in the composite cathode, such as stress-strain relation [17, 13], SE formability [18] and plasticity [19], significantly impact the development of internal stress within the CAM and SE particles and between them. As a result, mechanical stress/strain induce other impacts to the cathode, including the SE cracking, primary CAM particles fracturing, and disassembling of secondary CAM particles [14, 20]. These degradations accumulate during each cycle, leading to a rapid capacity fade and reduced cycle lifetime of the SSBs. [15]

To address these issues, researchers have proposed several strategies, including the use of zero-strain CAM materials [14, 15], radially aligned primary CAM particles [16] and hollow secondary CAM particles to create additional space for cathode deformation [21]. More recent studies also suggest using highly formable [22, 23] and pliable SE materials [18], which are related to the SE bulk modulus and hardness. High deformable and pliable SEs enable the maintaining of well-defined solid-solid interfaces with intimate contacts between CAM particles and SE particles during electro-chemical cycling. Additionally, external stack pressure has been considered to ensure the intimate CAM-SE contact, although it decreases the volumetric and gravimetric energy density with the usage of additional devices in the SSB cell [24].

Despite these strategies, the understanding and quantification of the mechanical behavior of the composite cathode during electrochemical cycling continue to be limited [25, 26]. Therefore, this paper aims to systematically investigate the mechanical behavior of the composite cathode during cycling. Three main mechanical defect indicators are defined and quantified: the composite cathode volume expansion at the end of discharge, the interfacial contact area loss between CAM and SE after one cycle, and the porosity development after one cycle. Other relevant secondary indicators, such as the interfacial detachment, the SE fracture and the SE tortuosity, are also discussed. The study suggests that SE materials with low elastic modulus are preferred to minimize these defects [16]. Furthermore, the paper proposes several strategies to alleviate mechanical failure through the interplay of factors such as the SE mechanical properties, external stack pressure, and CAM loading in the composite cathode. The work also provides predictive insights into the behavior of sulfide, oxide and polymer SE material types in the composite cathode.

**Methodology**

Composite solid-state cathodes in ASSBs comprise cathode active material (CAM), solid electrolyte (SE) particles, and carbon-based additives [9]. A mesoscale level model displaying CAM and SE particles in a cubic domain is shown in **Fig.** 1a [27]. To simulate the composite cathode, we employ a continuum Representative Volume Element (RVE), displayed in **Fig.** 1b as a half-3D model. To overcome the limitations of 3D simulations in terms of cost and accessibility, our two-part 3D model was transformed into a 2D model using axial- and plane-symmetric settings, as shown in **Fig.** 1c. Three main mechanical defect indicators were identified: volume expansion ($v_e$), interfacial contact area loss ($c_l$), and the porosity ($p_r$). *Volume expansion* ($v_e$) represents the percentage of the volume increment of the composite cathode at the end of discharge relative to its initial volume owing to CAM volume expansion, as indicated by dashed lines at the top edge of **Fig.** 1c. *Interfacial contact* area loss ($c_l$) represents the degree of interfacial detachment between the edges of CAM and SE following one cycle, as indicated by the dashed line at the CAM-SE interface. Porosity ($p_r$) represents the pores developed inside the composite cathode after one cycle. The developed pores can come from the SE or CAM mechanical deformations or the interface contact loss as indicated in **Fig.** 1c. Other secondary indicators such as *crack length $l_{SE}$* is also defined, representing the maximum lengths of propagated cracks within the SE after one cycle. All models were simulated with our house-developed code based on the open-source Finite Element framework MOOSE [28].

The composite cathode is assumed fully dense initially. This is achievable with high fabrication pressure or high temperature [29, 30]. The CAM particles evaluated in this work are secondary particles, having a size ranging between 1 to 10um, which are structures composed of thousands of primary particles that are randomly distributed and range in size from 10 to 100nm [27]. While the deformation of primary particles shows heterogeneity at the atomic scale [31, 12], it is assumed that the mesoscale expansion and contraction of the secondary CAM particles are homogenous as a result of random distribution of primary particles [32].

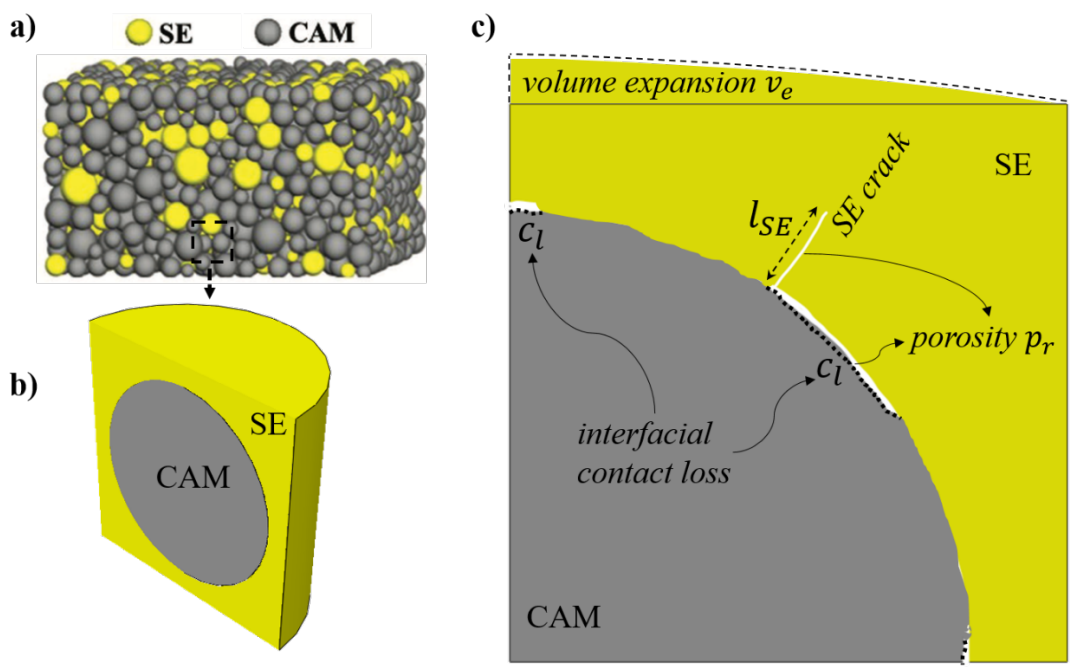

**Figure 1:** Schematic representative of the composite cathode at mesoscale. **(a)** Particles, **(b)** representative volume element, and **(c)** symmetrical 2D continuum model

Accurate mechanical properties of CAM and SE materials are crucial for simulating real experiments. Lithium Nickel Manganese Cobalt Oxide ($LiNi_{0.8}Co_{0.1}Mn_{0.1}O_2$, referred to as NMC or NCM), one of the most common cathode materials, has a reported Young's modulus of $E = 177.5\ (GPa)$ and Poisson's ratio of $v = 0.33$ [33, 34, 35], which was used for CAM particles in all simulations. Conversely, SE materials require consideration of a broader range of mechanical properties due to their more complex stress-strain behavior, including elastic, plastic, and fractural responses. Experiments were designed to obtain detailed stress-strain constitutive relations of SE materials, with details in the *supporting information (SI)-Section 1*. **Fig.** 2a shows the stress-strain relation of the SE pellet composed of $Li_6PS_5Cl$ (LPSCl) material under a single loading-unloading tensile test, where the black curve represents the measured original curve, and the blue curve is fitted to the bilinear isotropic hardening material model [36]. The initial non-linear section (<2%) of the original curve is ignored because of the pellet densification and experimental setup preparation. The tangent of the first linear segment corresponds to the elastic Young's modulus $E$, while that of the second segment corresponds to the plastic hardening tangent $E_t$. The elastic region is uniquely defined by the Young's modulus $E$ and yield strength $\sigma_y$, while the plastic-fracture region can be defined by the hardening tangent $E_t$, the ultimate strength $\sigma_u$, and plastic strain $\varepsilon_p$. These three plastic parameters are related through relation: $E_t = (\sigma_u - \sigma_y)/\varepsilon_p$. While numerous SE materials have been discovered over the last few decades [37, 38, 13, 39, 40, 17, 41], only limited mechanical properties have been investigated. For example, **Fig.** 2b presents an Ashby plot for elastic modulus $E$ and hardness $H_v$ of different SEs [13]. In the absence of some parameters, reasonable ranges are discussed to target an optimized combination of material properties. For example, a direct experimental measurement of the ultimate strength $\sigma_u$ of SE materials is absent in literature; however, the general relationship between the strength and the hardness of different materials has been investigated broadly in solid mechanics [42, 43] with the empirical relation: $H_v \approx 50 \sim 100\sigma_u$ for ceramics, similar material category as the SE materials.

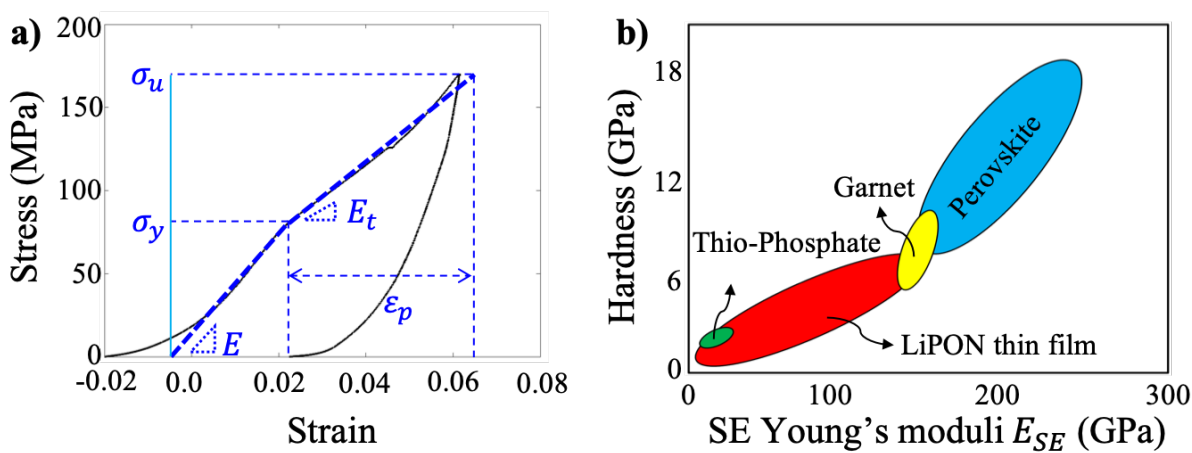

**Figure 2**: **(a)** Material constitutive behavior for LPSCl obtained from experiment overlaid with bi-linear elasto-plastic material model. **(b)** Ashby plot for different types of SEs.

Default values were used for each section unless otherwise stated explicitly. Isotropic CAM volume variation is set to 8.1% (2.6% in diameter) based on experimental and theoretical evidence [31]. The CAM diameter is 5 $\mu m$, and the SE volume is adjusted to maintain a 70:30 weight ratio of CAM to SE, with SE and CAM having mass densities of 1830 $kg/m^3$ (LPSCl) and 4750 $kg/m^3$ (NMC) respectively [33, 40]. A stack pressure of 5 MPa is applied to the top boundary, and the right boundary is fixed due to the lateral confinement.

**Figure** 3 presents snapshots of the composite cathode's deformation at different (dis)charging states, along with the accompanying defect indicators. A more detailed simulation video is provided in *SI-Movie 1*. During discharge (t=0 to 0.50 h), the CAM particles expand and exert stress on the interface, causing volume expansion as well as the propagation of SE cracks. The maximum volume expansion of the composite cathode ($v_e$) is determined at the end of discharge. On the other hand, during charge (t=0.5 h to 1 h), CAM contracts and returns to its original condition, while the SE part, owing to residual plastic strains in the material, is unable to return to its initial position, resulting in the development of interfacial contact loss ($c_l$) and porosity ($p_r$). The permanent volume variation in composite cathode after a single discharge-charge cycle can be determined from the cathode's final configuration (bottom-right subplot), which differs from the cathode volume expansion $v_e$ defined at the end of discharge (top-right subplot). Likewise, the cathode porosity after one cycle can be extracted from developed pores due to contact loss and SE crack. For instance, at point *A* in **Fig.** 3, the CAM and SE part adhere up to $t = 0.7h$, but are subsequently separated with point *B* propagating along the interface. The contact loss at the end of charge ($t = 1h$) is extracted from the ratio of the length of *AB* to the total interface AO ($c_i = \overline{AB}/\overline{AO}$). The specific location at the interface is defined as "contact loss" when its detached distance is larger than 10nm, which is the model's numerical resolution [44]. The final porosity $p_r$

in the composite cathode following one electrochemical cycling is calculated by the areal integration of the contact loss interface $\overline{AB}$ and the SE crack $l_{SE}$.

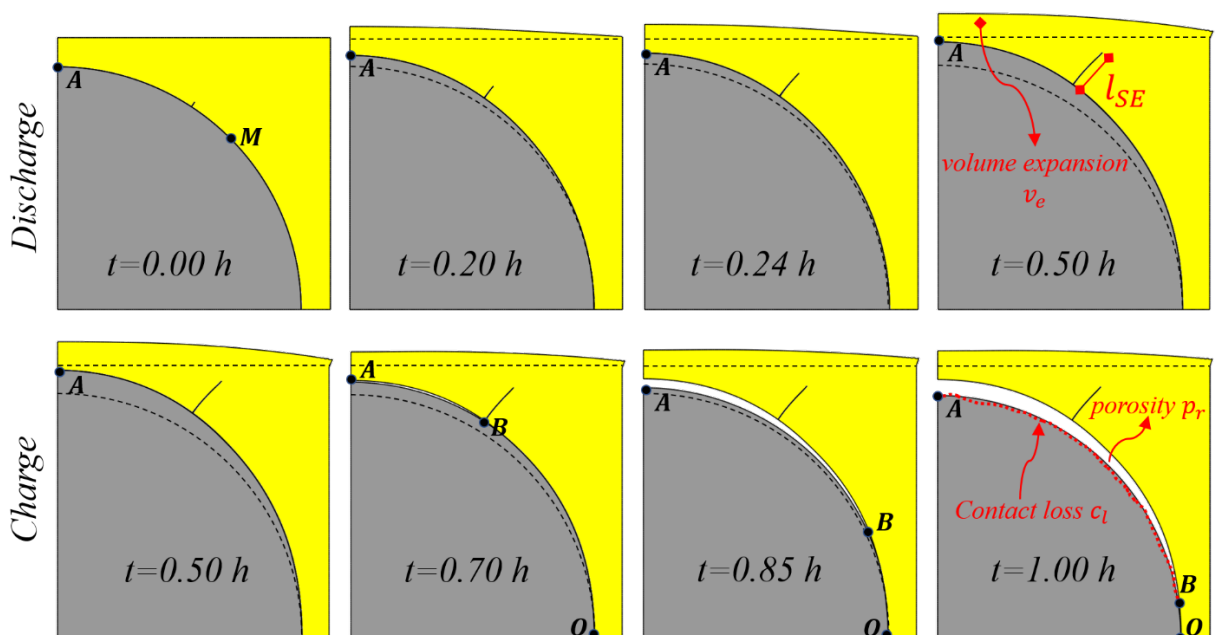

**Figure 3:** Morphology evolution (from left to right) of the composite cathode (discharge at top and charge at the bottom). Dashed lines represent original boundary of CAM and SE. Point $B$ propagates from point $A$ to point $O$ along the CAM/SE interface. Point $M$ locates at the middle of the interface for later usage.

**Results**

**Effect of SE material properties**

During the electrochemical cycling of the SSB cell, the SE material in the composite cathode undergoes an elasto-plasto-fracture response. Different SE materials, characterized by their distinct material properties, causing varying response of mechanical defects in the composite cathode. The effects of the elastic and plastic properties of SE materials are discussed in **Figure** 4 and **Figure** 5 respectively. Additional data is provided in Table S1 in the *SI-Section 2*.

In general, the mechanical defect indicators, including volume expansion, interfacial contact loss, and porosity, all increase as the SE material becomes stiffer elastically and plastically. This behavior is illustrated in **Fig.** 4a and **Fig.** 5a for volume expansion, in **Fig.** 4b and **Fig.** 5b for interfacial contact loss, and in **Fig.** 4c and **Fig.** 5c for porosity. The stiffness of SE materials are represented by their respective properties on the x-axis: the SE Young's modulus $E$ in **Figure** 4 and the SE plastic tangent $E_t$ in **Figure** 5. Specifically, the mechanical defects grow with an increase of the SE material's yield strength $\sigma_y$ in **Figure** 4 and ultimate strength $\sigma_u$ in **Figure** 5.

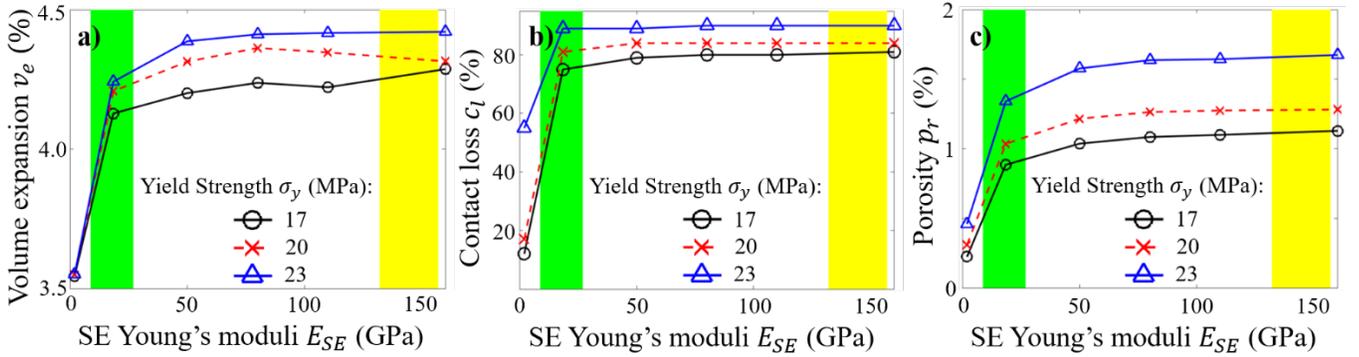

**Figure 4:** Effect of elastic properties of SE material on the defect indicators. x-axis represents the SE Young's moduli and different curves represents different SE yield strengths. The green region is the Young's modules ($E_{SE}$) for sulfide-type SE and the yellow region is for oxide-type SE. Choice of SE material with smaller $E_{SE}$ and $\sigma_y$ can minimizing the mechanical defects.

The elastic region can be uniquely defined by two parameters: the Young's modulus $E$ and the yield strength $\sigma_y$. **Figure** 4 illustrates the impacts of these two elastic parameters on the three defect indicators, while setting the plastic parameters constant values ($E_t = 1 GPa, \varepsilon_p = 0.004$). It is clear that a softer SE material such as sulfide-type SE (green region) causes much smaller mechanical defects in the composite cathode than stiffer SE material such as oxide-type (yellow region) SE after one cycle. When the Young's modulus $E$ is less than 20 $GPa$ in the green region, all mechanical defects decrease significantly. As indicated by the first data points of all three mechanical defects in **Fig.** 4a-4c, the value of these defects under different yield strength $\sigma_y$ converges when $E$ is small enough (~2$GPa$). Further analysis of the stress/strain state in Figure SI-2 of *SI-Section 3* indicates that the strain state in the SE part decreases significantly when $E_{SE}$ is small, despite the same volume expansion of 8.1% from the CAM part. The small development in strain dominates the cathode deformation throughout the entire electrochemical cycle. Thus, SE with a low Young's modulus consistently demonstrates small mechanical defect indicators, regardless of yield strength $\sigma_y$. However, when the SE material becomes stiffer with a higher Young's modulus $E$, the impact of the yield strength $\sigma_y$ becomes increasingly significant. The stress/strain analysis in Figure SI-2 of *SI-Section 3* shows that a much higher stress and strain state are developed in the composite cathode when $E_{SE}$ exceeds 100 $GPa$. For a given high $E_{SE}$ value such as in the yellow region, the stress increases rapidly and approaches to the yield strength $\sigma_y$. SE material with smaller yield strength reaches plasticity earlier than that with higher yield strength (as indicated in **Fig.** 2a), leading to higher degree of plastic flow into volumetric pores and interfacial contact loss area developed during the cycle. Therefore, if SE materials with high Young's modulus are to be used in the composite cathode, smaller values of the SE yield strength (or hardness) are preferred to enable some degree of plastic flow of the SE material and minimize the mechanical defects in the composite cathode.

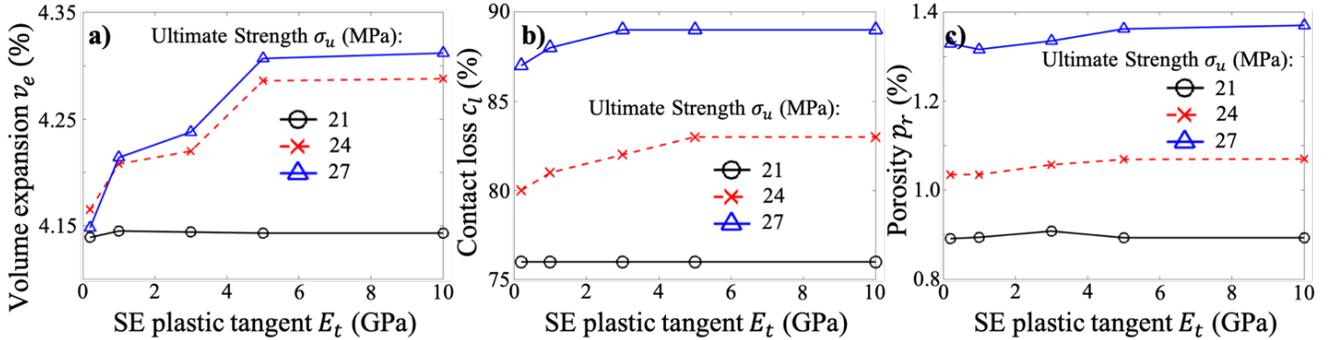

**Figure 5:** Effect of plastic properties of SE material on the defect indicators. x-axis represents the SE plastic tangent and different curves represents different SE ultimate strengths. Choice of SE material with smaller $\sigma_u$ can minimizing the mechanical defects.

The plastic region can be uniquely defined by the hardening tangent $E_t$ and the ultimate strength $\sigma_u$. The plastic strain $\varepsilon_p$ can be determined through the relation: $\varepsilon_p = (\sigma_u - \sigma_y)/E_t$. **Figure** 5 demonstrates the impact of the two plastic parameters on the three defect indicators with elastic parameters of Young's modulus $E = 18.5 GPa$ and yield strength $\sigma_y = 20 MPa$ (LPSCl, one sulfide-type SE). Similar to the results of the elastic response, SE material with smaller plastic tangent and ultimate strength exhibit lower mechanical defects in the cathode.

The volume expansion $v_e$ in **Fig. 5a** is strongly influenced by both the plastic tangent $E_t$ and the ultimate strength $\sigma_u$. In contrast, the interfacial contact area loss $c_l$ in **Fig. 5b** and the porosity $p_r$ in **Fig. 5c** are less sensitive to the plastic tangent but more to the ultimate strength. Thus, the ultimate strength $\sigma_u$ is a more efficient plastic parameter for minimizing the mechanical defects resulting from the plastic response of the SE material. Notably, inorganic electrolytes behave like brittle ceramics with little or zero plastic deformations ($\sigma_u \approx \sigma_y$ according to **Fig. 2a**) [45], whereas organic electrolytes behave like ductile polymers with significant plastic deformations ($\sigma_u \gg \sigma_y$ [46]. Therefore, to minimize mechanical defects due to the SE material's plastic response, one can use inorganic SEs due to their brittle nature or choose organic polymer SEs with lower ultimate strength.

**Effect of external stack pressure**

Previously reports have suggested that stack pressure is a suitable approach to mitigate cathode volume variation and maintain intimate interfacial contact [18, 47, 48]. Here, further quantification is provided for the effect of external stack pressure when different SE materials are used in the composite cathode. **Figure** 6 depicts the evolution of the three defect indicators (i.e. volume change $v_e$ in **Fig. 6a**, contact loss $c_l$ in **Fig. 6b**, and porosity $p_r$ in **Fig. 6c**) alongside Young's moduli (E$_{SE}$) for different SE materials. The four curves in each plot represent the results of applying four different stack pressures (1, 3, 5, and 7 MPa) to the top boundary. Additional data is provided in Table SI-1 in the *SI-Section 2*.

In general, increasing stack pressure results in reduced cathode degradation, as reflected by decreasing trends of all three mechanical defect indicators under stack pressure from 1 MPa to 7 MPa. However, softer SE materials, marked in green as sulfide-type SE, are more affected by external stack pressure compared to their stiffer counterparts, marked in yellow as oxide-type SE. For instance, for LPSCl (one sulfide-type SE with $E_{SE} = 18.5$ GPa), the volume expansion, contact area loss and porosity reduce by 29.3%, 77%, and 96.4%, respectively when the applied stack pressure increases from 1 MPa to 7 MPa which is consistent with the experimental reports [18, 49]. However, these values are 21.2%, 27%, and 83.2%, respectively for LLZO (one oxide-type SE with $E_{SE} = 155$ GPa) under the same stack pressure increases. Moreover, the suppression of mechanical defects is highly nonlinear to the applied stack pressure for different SE Young's modulus values. For example, a stack pressure of 5MPa considerably decrease all mechanical defects compared to the value of 3MPa only for SE materials with Young's modulus less than 50GPa. The 5MPa stack pressure does not decrease the porosity defect in **Fig.** 6c when $E_{SE}$ is larger than 80GPa, and makes no change to the volume expansion in **Fig.** 6a and contact loss in **Fig.** 6b when $E_{SE}$ reaches 155GPa. Therefore, stiffer SE materials with higher Young's modulus not only become less sensitive to the applied stack pressure but also require higher values to suppress the mechanical defects.

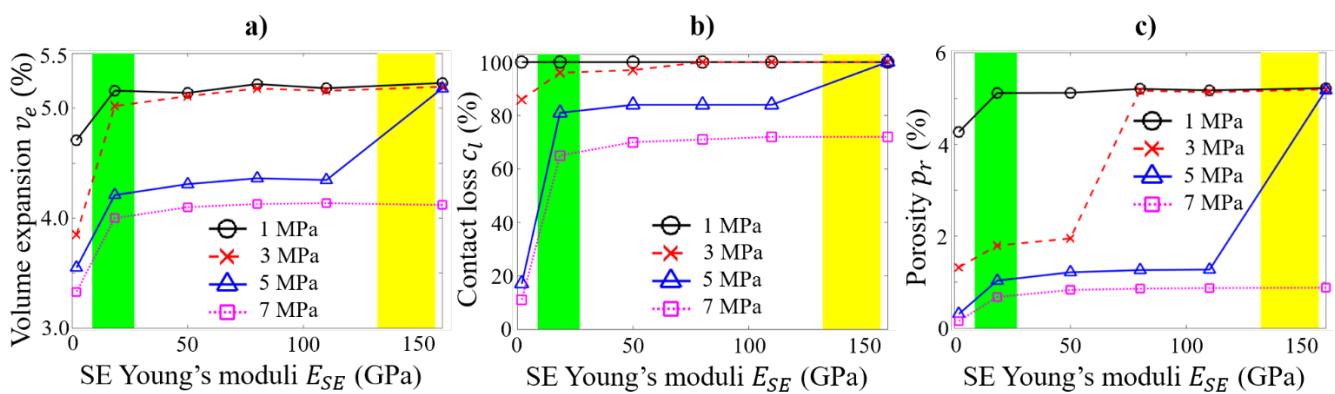

**Figure 6:** Stack pressure effect on three defect indicators: **a)** maximum cathode volume change $v_e$; **b)** interfacial contact between NMC and SE $c_l$; **c)** cathode porosity $p_r$. The green region is the Young's modules ($E_{SE}$) for sulfide-type SE and the yellow region is for oxide-type SE. Higher stack pressure decreases mechanical defects with (1) nonlinear behavior and (2) more impact on softer SE materials (lower $E_{SE}$).

Notably, although high stack pressure is helpful to decrease mechanical defects in the composite cathode, it also causes enhanced dendrite growth when goes very high due to the fracture of the SE conductor [50, 51].

**Effect of CAM state of discharge and CAM loading**

The volume change of the secondary cathode active material (CAM) is determined by the state of discharge (SOD) of each primary CAM particle. Previous studies have demonstrated that the fully delithiated NMC811 ($Ni_{0.8}Mn_{0.1}Co_{0.1}O_2$) can expand by 12.5% when fully discharged to the lithiated state ($LiNi_{0.8}Mn_{0.1}Co_{0.1}O_2$) [52, 31]. Additionally, the composite cathode's volume expansion $v_e$ is influenced by the CAM loading (the weight ratio of CAM). **Figure** 7 depicts the impact of both the SOD and the CAM loading on all the three mechanical defects. For the purpose of this study, it is assumed that all CAM particles can be fully utilized through structure optimization [27]. The sulfide material LPSCl are used as SE, with detailed description in Table SI-1 in the *SI-Section 2*.

**Fig.** 7a depicts the interfacial detachment of point *M*, as defined in **Fig.** 3, after a single cycle under varying SOD values. SOD is defined as 100% when the primary NMC particles are discharged to the lithiated state, resulting in a volume change of 8.1%. However, theoretical predictions suggest that it could reach up to 12.5% [52, 31]; therefore, higher SOD values up to 150% are also considered in this study. Reducing the SOD leads to a slower increase in the displacement of point *M* during discharge, resulting in a smaller peak value at the end of discharge (t= 0.5 hour). When SOD is less than 80%, residual displacement of point *M* is small (<20nm) at the end of charge (t=

1 hour), indicating that the composite cathode has recovered without serious permanent volume change (<3%). The SE part has slightly entered plasticity but is still primarily in elastic after one cycle. When SOD is greater than 120%, the residual displacement becomes significantly large (> 60nm), and more of the SE component enters plasticity until it is fully plasticized.

Fig. 7b presents a more quantitative analysis with the evolution of the three defect indicators as a function of different SOD. The value of $v_e$ represents the same moment when point $M$ reaches its peak displacement (t= 0.5 hour). The blue dashed line represents the volume expansion of only the CAM part at the same time. The difference between the two curves is the volume contraction of the SE part, from both elastic and plastic deformation. When SOD is great than 100%, the difference between the two lines increases, indicating an increase in the plastic response of the SE part. This trend is more clearly illustrated by the interfacial contact loss $c_l$ in the middle subplot of **Fig.** 7b. It increases slowly when SOD is less than 80%, indicating major SE elasticity and minor SE plasticity. It surges up quickly when SOD is greater than 100%, indicating major SE plasticity. Regarding the porosity development in the bottom subplot of **Fig.** 7b, small value (<0.5%) is developed before SOD is less than 80% due to elastic response. Overall, higher SOD is expected for higher cathode capacity while larger mechanical defects are unexpectedly developed.

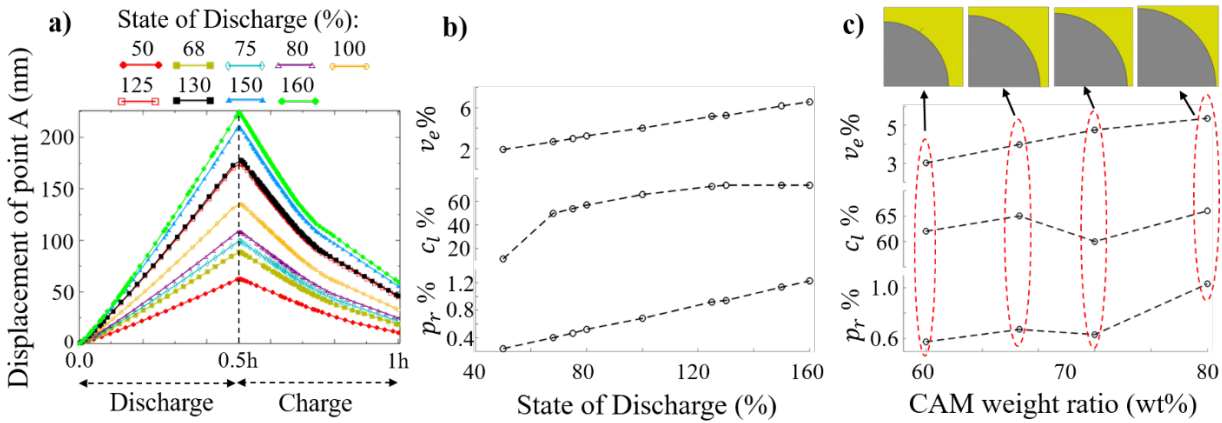

**Figure 7: (a)** the cathode volume change after one single cycle at different state of discharge. The relation between the three defect indicators (volume expansions $v_e$, interfacial contact area loss $c_l$ and the cathode

porosity $p_r$) as a function of the **(b)** state of discharge, and **(c)** the weight ratio of CAM to SE. Lower state of discharge and weight ratio decreases mechanical defects.

The effect of CAM loading on the three defect indicators is shown in **Fig.** 7c. Four CAM weight ratio ranging from 60 wt.% to 80 wt.% are considered, with the given SOD of 100% (8.1% CAM expansion). The SE part occupies less volume in the composite when the CAM weight ratio increases, as illustrated in the top subplots of **Fig.** 7c. The cathode volume expansion $v_e$ increases from 3% to 5% as the CAM weight ratio increases from 60 wt.% to 80 wt.%, and approaches to the CAM volume expansion (8.1%) if higher CAM loading is used in the composite. The porosity $p_r$ stays stable (0.6%) when the CAM loading is less than 75 wt.%, but doubles its value to >1% if higher CAM loading (80 wt.%) is tried. Interestingly, the interfacial contact loss $c_l$ decreases as the CAM loading increases. This is because less SE material exists in the composite to buffer the CAM volume expansion at the end of discharge, resulting in less overall SE plastic deformation after one cycle. As more SE materials deform elastically and are back to the original location, less contact loss is developed at the SE/CAM interface.

## Discussion

High CAM loading is desired in the composite cathode of the SSB cell, in order to improve the cathode energy density [53, 54]. However, earlier work shows that too high CAM loading can causes decreased utilization of CAM particles due to insufficient ionic diffusing pathways between the CAM material and the limited SE material [27]. Here, we further show that high CAM loading may also cause serious development of mechanical defects in the composite cathode during the electrochemical cycling (**Figure 7**). To minimize the mechanical defects while targeting high cathode loading, suitable experimental conditions (**Figure 6**) need to be considered together with the SE and CAM materials (**Figure 4-5**) selected for the composite cathode. A design principle that correlates the experimental setup (such as CAM loading and stack pressure) and the internal mechanical degradations (such as volume change, contact loss, and porosity development) is practically useful for the guidance of optimizing the performance of SSB cell.

**Figure 8** provides such a design principle when NMC is used as CAM material and different SE materials are used in the composite cathode, include the oxide-type LLZO (**Fig.8a**), the sulfide-type LPSCl (**Fig.8b**), and the predicted SE material with desired mechanical properties (**Fig.8c**). Additional data is provided in Table SI-2 in the *SI-Section 2*. The contour represents the "degree-of-defect", which is based on the results of three mechanical defect indicators (cathode volume change $v_e$, interfacial contact area loss $c_l$, and the porosity $p_r$). Three threshold values are defined with $v_e = 3\%$, $c_l = 50\%$, and $p_r = 1\%$ for the three mechanical defect indicators respectively. The degree-of-defect is zero ("defect-free") when all the developed mechanical defects in the cathode are smaller than their corresponding threshold values, which is colored in green in **Figure 8b-c**. The degree-of-defect are 1 or 2 or 3 when one or two or three mechanical defects are larger than their threshold values, which are colored in yellow or orange or red in **Figure 8**, respectively. The x-axis is the applied stack pressure to the composite cathode and the y-axis is the CAM volume ratio used in the composite cathode. The CAM loading (weight ratio) can be obtained from the CAM volume ratio based on the mass density of the selected CAM and SE materials ($\rho_{NMC} = 4750 \ kg/m^3$, $\rho_{LPCl} = 1830 \ kg/m^3$, and $\rho_{LLZO} = 5100 \ kg/m^3$). Notably, the threshold values used here are based on experimental observations of good SSB performance from literature [55]. These values can be defined smaller as more strict "defect-free" criteria for the design principle of the composite cathode, or defined larger as less strict "defect-free" criteria for the design principle.

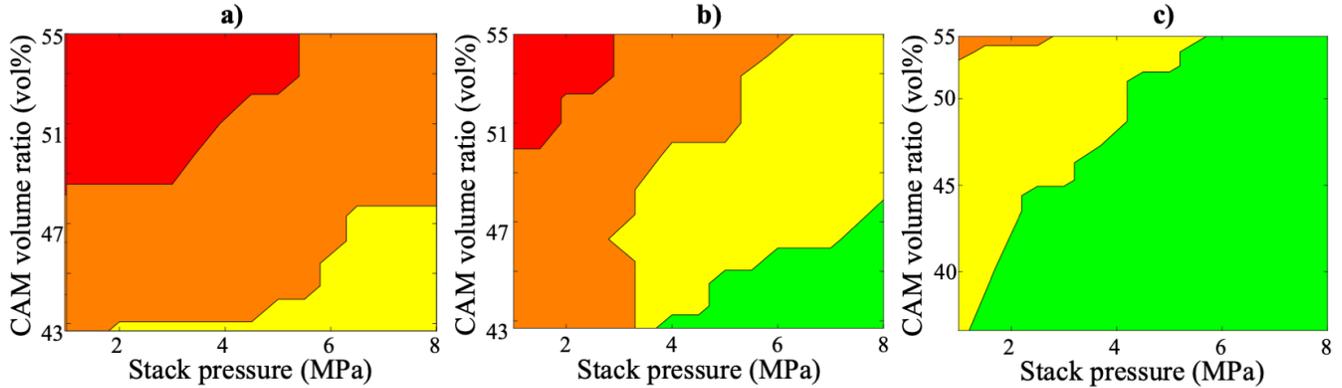

**Figure 8:** Design principle for the composite cathode when three different SE materials are used with NMC as the cathode active material: **a)** LLZO **b)** LPSCl, and **c)** SE material with low Young's modulus and hardness. The contour in each plot represents the degree of defect in the cathode with values 0,1,2,3, corresponding to the situation when 0,1,2,3 mechanical defects are larger than the threshold values ($v_e = 3\%$, $c_l = 50\%$, and $p_r = 1\%$).

The result in **Fig.8a** shows there is no "defect-free" (green color) status if using LLZO as SE material in the composite cathode, even at high stack pressure up to 8MPa and low CAM loading at 43%. Large interfacial contact area loss (>50%) is developed first (yellow region), followed by the increasing porosity (>1%, orange region) and the volume change (>3%, red region). This is due to the high intrinsic stiffness and hardness of LLZO (E=150GPa and H=6GPa), as discussed in **Figure 4**. Therefore, it is hard to reach high CAM volume ratio if using LLZO as SE material in the composite cathode due to the considerable development of mechanical defects in the cathode. Considering the mass densities of oxide-type SE materials are usually high, the corresponding CAM loading would be even lower (for example, CAM loading ranges from 40 wt% to 52 wt% for the LLZO case in **Fig.8a**, compared to the range from 65 wt% to 75 wt% for the LPSCl case in **Fig.8b**). This leads to the conclusion that oxide-type SE materials are not good candidates for the composite cathode due to low volumetric and gravimetric energy density.

On the contrary, the composite cathode with LPSCl (E=18.5GPa and H=2GPa) as SE material in **Fig.8b** shows much less mechanical defects. The defect-free state can be reached at low CAM loading (45 vol%) with high stack pressure (>5MPa). However, mechanical defects are developed gradually in the cathode when increasing the CAM loading or decreasing the stack pressure. For example, if targeting on more than 50 vol% of CAM material, a stack pressure larger than 8MPa is needed to maintain the defect-free cathode. This requirement complicates the engineering design and decreases the energy density of the battery component. Therefore, sulfide-type SE materials may be used in the composite cathode to suppress the development of mechanical defects but better SE materials with more desirable mechanical properties are needed. **Fig.8c** shows the degree of defects when a theoretical SE material with low Young's modulus and hardness (E=1GPa and H=0.2GPa) is used in the cathode. A much smaller stack pressure (5MPa) is needed for the composite cathode even the CAM volumetric ratio goes as high as 53 vol%. For low degree of defect (yellow region) where only the contact area loss is larger than threshold value (>50%), the CAM volumetric loading can be higher than 55 vol% at sub-MPa stack pressure. Therefore, it is important to screen the correct SE material with the desirable mechanical properties.

However, the current criteria for screening SE materials are mainly focusing on their electrochemical properties, such as high ionic conductivity [37, 56] and excellent (electro)chemical stabilities [57]. From the conclusion in **Fig.8c**, it is clear the mechanical properties of SE need to be considered when selecting the proper SE material for the composite cathode. For example, small Young's modulus (**Figure 4**) and low ultimate strength (**Figure 5**) of SE materials are preferred to minimize these mechanical defects. New SE selecting criteria should be designed based on both the electrochemical and the mechanical properties.

This work focuses on the major mechanical properties in the composite cathode by considering as much mechanical aspects as possible; however, there are still other mechanical factors need to be explored and added into the design principle as proposed above. For example, SE/CAM fracture is frequently observed in the composite cathode during cycling. This SE fracture can be detrimental to the cathode since it leads to higher tortuosity and frequent inter-granular transport of Li-ions between SE particles. Eventually the internal resistance of the composite cathode will be increased. A preliminary crack analysis is described in *SI-Section 4* and *SI-Movie 2*, and more detailed work needed in the future. Furthermore, it has been shown that the interfacial adhesion between the SE and the CAM particles may also play important role for the interfacial contact [58, 59]. There is a pressing need for carefully designed experiments that can explicitly extract the effect of mechanical properties of the composite cathode.

**Conclusion**

This theoretical study on the composite cathode of SSB indicates that the mechanical defects including volume expansion, interfacial contact loss, porosity, and cracks are a function of different parameters such as stack pressure, volume change of CAM part, weight ratio (CAM: SE), constitutive material properties, and boundary conditions. In general, increasing the elastic modulus of SE material increases all mechanical defects. Our results highlight the importance of SE material with appropriate constitutive behaviors. For example, low elastic modulus materials like Thio-Phosphate type with low yield/ultimate stress seem to be promising. As for external pressure, while applying high pressures considerably decreases all adverse mechanical degradations, dendrite growth may happen when the value is too high. This implies the dominant role of material properties in mechanics of composite solid-state composite. In conclusion, this study not only offer predictive insights towards the mechanic-based response of some existing sulfide, oxide, and polymer SE material type, but also open windows on proper design strategies/guidelines for composite solid-state cathodes by fine-tuning material properties, optimizing stack pressure, and other strategies.

**Acknowledgment**

This work was supported by the Samsung Advanced Institute of Technology. This research used the SPORC computational cluster resource provided by the IT Division at Rochester Institute of Technology. The authors also gratefully acknowledge Dr. Hailong Chen from Georgia Institute of Technology, Dr. Chen Ling from Toyota for insightful discussions.